\documentclass[reprint, superscriptaddress, twocolumn, amsmath, amssymb, aps, prb]{revtex4-2}
\usepackage{graphicx}% Include figure files
\usepackage{dcolumn}% Ali$G_N$ table columns on decimal point
\usepackage{bm}% bold math
\usepackage{placeins}
\usepackage{appendix}
\usepackage{verbatim}
\usepackage[usenames,dvipsnames]{xcolor}
 \usepackage{amsmath}
 \usepackage{amsfonts}
 \usepackage{amssymb}
 \usepackage[colorlinks=true,citecolor=blue,linkcolor=red]{hyperref}

 \usepackage{bbold}     % for \mathbb{1} as a nice unit matrix symbol
 \usepackage[makeroom]{cancel}  % crossed terms in tables
 \usepackage{multirow}    % merge cells vertically in tables
 \usepackage[normalem]{ulem}        % strike-through text
 \usepackage{array}
 \usepackage{pdfpages}
\makeatletter
 \AtBeginDocument{\let\LS@rot\@undefined}
 \makeatother

\begin{document}

%\linenumbers

\title{Proximity effect in PbTe-Pb hybrid nanowire Josephson junctions}

\author{Zitong Zhang}
 \email{equal contribution}
\affiliation{State Key Laboratory of Low Dimensional Quantum Physics, Department of Physics, Tsinghua University, Beijing 100084, China}

\author{Wenyu Song}
\email{equal contribution}
\affiliation{State Key Laboratory of Low Dimensional Quantum Physics, Department of Physics, Tsinghua University, Beijing 100084, China}

\author{Yichun Gao}
 \email{equal contribution}
\affiliation{State Key Laboratory of Low Dimensional Quantum Physics, Department of Physics, Tsinghua University, Beijing 100084, China}

\author{Yuhao Wang}
 \email{equal contribution}
\affiliation{State Key Laboratory of Low Dimensional Quantum Physics, Department of Physics, Tsinghua University, Beijing 100084, China}

\author{Zehao Yu}
\affiliation{State Key Laboratory of Low Dimensional Quantum Physics, Department of Physics, Tsinghua University, Beijing 100084, China}

\author{Shuai Yang}
\affiliation{Beijing Academy of Quantum Information Sciences, Beijing 100193, China}

\author{Yuying Jiang}
\affiliation{State Key Laboratory of Low Dimensional Quantum Physics, Department of Physics, Tsinghua University, Beijing 100084, China}

\author{Wentao Miao}
\affiliation{State Key Laboratory of Low Dimensional Quantum Physics, Department of Physics, Tsinghua University, Beijing 100084, China}

\author{Ruidong Li}
\affiliation{State Key Laboratory of Low Dimensional Quantum Physics, Department of Physics, Tsinghua University, Beijing 100084, China}

\author{Fangting Chen}
\affiliation{State Key Laboratory of Low Dimensional Quantum Physics, Department of Physics, Tsinghua University, Beijing 100084, China}

\author{Zuhan Geng}
\affiliation{State Key Laboratory of Low Dimensional Quantum Physics, Department of Physics, Tsinghua University, Beijing 100084, China}

\author{Qinghua Zhang}
\affiliation{Institute of Physics, Chinese Academy of Sciences, Beijing 100190, China}

\author{Fanqi Meng}
\affiliation{School of Materials Science and Engineering, Tsinghua University, Beijing 100084, China}

\author{Ting Lin}
\affiliation{Institute of Physics, Chinese Academy of Sciences, Beijing 100190, China}

\author{Lin Gu}
\affiliation{School of Materials Science and Engineering, Tsinghua University, Beijing 100084, China}

\author{Kejing Zhu}
\affiliation{Beijing Academy of Quantum Information Sciences, Beijing 100193, China}

\author{Yunyi Zang}
\affiliation{Beijing Academy of Quantum Information Sciences, Beijing 100193, China}

\author{Lin Li}
\affiliation{Beijing Academy of Quantum Information Sciences, Beijing 100193, China}

\author{Runan Shang}
\affiliation{Beijing Academy of Quantum Information Sciences, Beijing 100193, China}

\author{Xiao Feng}
\affiliation{State Key Laboratory of Low Dimensional Quantum Physics, Department of Physics, Tsinghua University, Beijing 100084, China}
\affiliation{Beijing Academy of Quantum Information Sciences, Beijing 100193, China}
\affiliation{Frontier Science Center for Quantum Information, Beijing 100084, China}

\author{Qi-Kun Xue}
\affiliation{State Key Laboratory of Low Dimensional Quantum Physics, Department of Physics, Tsinghua University, Beijing 100084, China}
\affiliation{Beijing Academy of Quantum Information Sciences, Beijing 100193, China}
\affiliation{Frontier Science Center for Quantum Information, Beijing 100084, China}
\affiliation{Southern University of Science and Technology, Shenzhen 518055, China}

\author{Ke He}
\email{kehe@tsinghua.edu.cn}
\affiliation{State Key Laboratory of Low Dimensional Quantum Physics, Department of Physics, Tsinghua University, Beijing 100084, China}
\affiliation{Beijing Academy of Quantum Information Sciences, Beijing 100193, China}
\affiliation{Frontier Science Center for Quantum Information, Beijing 100084, China}

\author{Hao Zhang}
\email{hzquantum@mail.tsinghua.edu.cn}
\affiliation{State Key Laboratory of Low Dimensional Quantum Physics, Department of Physics, Tsinghua University, Beijing 100084, China}
\affiliation{Beijing Academy of Quantum Information Sciences, Beijing 100193, China}
\affiliation{Frontier Science Center for Quantum Information, Beijing 100084, China}

%\date{\today}

\begin{abstract}

Semiconductor-superconductor hybrid nanowires are a leading material platform for the realization of Majorana zero modes. The semiconductors in previous studies are dominantly InAs or InSb. In this work, we show the induced superconductivity in a PbTe nanowire epitaxially coupled to a superconductor Pb. The Josephson junction device based on this hybrid reveals a gate-tunable supercurrent in the open regime and a hard superconducting gap in the tunneling regime. By demonstrating the superconducting proximity effect, our result can enable Majorana searches and other applications like gate-tunable qubits in a new semiconductor system.

\end{abstract}

\maketitle

A semiconductor nanowire coupled to a superconductor is an intriguing quantum system owing to the proximity effect. One such example is the gate-tunable Josephson junction \cite{Leo_Supercurrent} which plays a key role in the gatemon superconducting qubit \cite{2015_PRL_gatemon}. Moreover, the interplay between the strong spin-orbit coupling in the semiconductor and the Zeeman energy may lead to topological phases hosting Majorana zero modes \cite{Lutchyn2010, Oreg2010, Prada2020, NextSteps}. InAs and InSb nanowires are the semiconductors commonly used in those studies due to several practical reasons, e.g., the well established state-of-art device fabrication and control \cite{Kammhuber2016, Gul2017, Zhang2017Ballistic}, epitaxial growth of superconductors \cite{Chang2015, Roy, PanCPL} and the strong spin-orbit interaction \cite{Ilse_2015, Jouri2019}. These advantages are all crucial and indeed have enabled tremendous experimental progress on possible Majorana signatures \cite{Mourik, Deng2016, Albrecht, Gul2018, Zhang2021, Song2022, WangZhaoyu}. The current roadblock is device disorder which has to be improved first before further progress \cite{Prada2012, Patrick_Lee_disorder_2012, Liu2017, DEL-Disorder2018, CaoZhanPRL, Loss2018ABS, GoodBadUgly, DasSarma2021Disorder, DasSarma_estimate, DasSarma_random_matrix, Tudor2021Disorder, SauQuality}. 

To overcome this challenge, PbTe nanowires have recently been proposed as a potentially better candidate \cite{CaoZhanPbTe}. The hope is that the large dielectric constant ($\sim$ 1350) in PbTe can significantly screen charge disorder. Moreover, growing PbTe on a lattice-matched substrate, CdTe, can further reduce the substrate disorder \cite{Jiangyuying}. Finally, capping the PbTe with CdTe can push the surface disorder away from the core region of the device. Quickly, experimental efforts \cite{Jiangyuying, Erik_growth_PbTe, PbTe_AB, Fabrizio_PbTe, Erik_PbTe_SAG} have been carried out on the growth of PbTe nanowires with transport characterizations on the field effect mobility, weak anti-localization, Aharonov-Bohm oscillations and quantum dots. So far, the key question of whether superconducting proximity effect in PbTe nanowires exists is still pending. Here, we demonstrate the induced superconductivity in a PbTe nanowire coupled to a superconductor Pb. Two hallmark transport signatures can be revealed in a Josephson junction (JJ) device based on this hybrid: a gate-tunable supercurrent in the open regime and a hard superconducting gap in the tunneling regime. Our result may open the door to a new wave of Majorana studies and also other quantum devices, e.g. hybrid qubits \cite{2015_PRL_gatemon}.

\begin{figure}[b]
\includegraphics[width=\columnwidth]{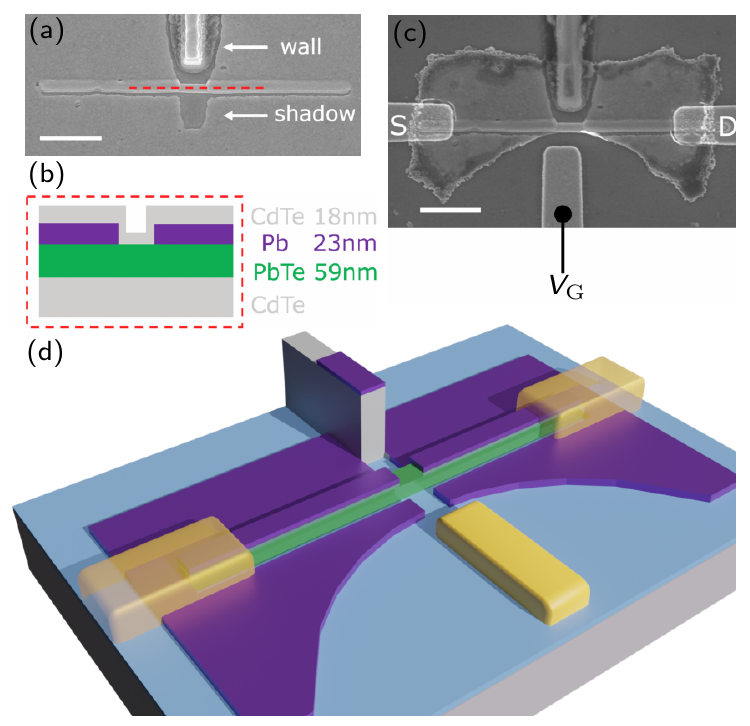}
\centering
\caption{ (a) SEM of a PbTe-Pb nanowire. Scale bar, 1 $\mu$m. (b) The JJ schematic, cut at the red dashed line in (a). Layer thickness on the right.  (c) SEM of device A. Scale bar, 1 $\mu$m. Contacts and gate, Ti/Au (10/65 nm). (d) Device schematic.  }
\label{fig1}
\end{figure}

Figure 1 shows the scanning electron micrograph (SEM) of a PbTe-Pb nanowire and its device schematic. The nanowire growth is similar to that in Ref. \cite{Jiangyuying} with minor modifications. The CdTe substrate was first covered by Si$_x$N$_y$. Nanowire-shape trenches were defined by etching Si$_x$N$_y$ using reactive ion etching. The chip was then loaded into a molecular beam epitaxy chamber for the nanowire growth. The chip was cleaned with Ar treatment, annealed at 240.6 $^{\circ}$C, and then followed by selective area growth of the CdTe buffer. This procedure ensures that the PbTe nanowire is spatially separated from the disorder generated during the substrate cleaning. PbTe was then grown, followed by the Pb film deposition at a tilted angle without breaking vacuum. The standing wall shadowed part of the PbTe nanowire, forming a JJ. The entire chip was then capped by CdTe to prevent oxidation. Figure 1(a) is the SEM after growth and Fig. 1(b) is the JJ layer structure (not in scale).

For device fabrication, most of the Pb film on the substrate was etched using Ar ion milling to prevent a short circuit. Source/drain (S/D) electrodes and a side gate were fabricated by evaporating Ti/Au. Ar-plasma etching was performed before the evaporation to remove the CdTe capping in the contact regions for ohmic contacts. Figures 1(c) and 1(d) show the SEM and a 3D schematic of the final device. For the growth and fabrication details, see the method section in the supplementary materials (SM). The device was measured in a dilution fridge at a base temperature $T \sim$15 mK using the standard two-terminal set-up.

\begin{figure}[b]
\includegraphics[width=\columnwidth]{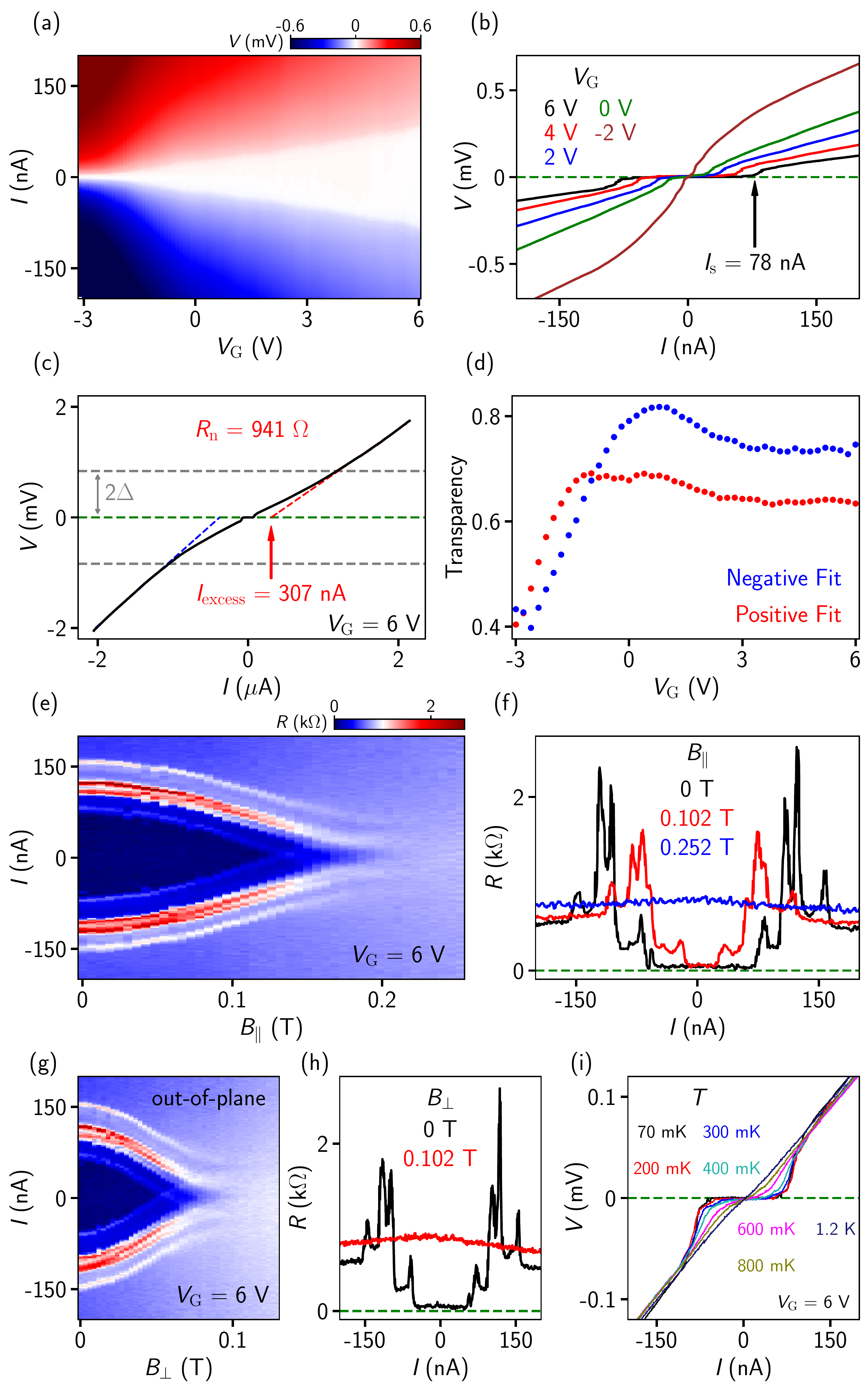}
\centering
\caption{Supercurrent in the open regime of Device A. (a) $I$-$V$ characteristic as a function of $V_{\text{G}}$. $B$ = 0 T. (b) Line cuts from (a). (c) $I_{\text{excess}}$ and $R_{\text{n}}$ estimations of an example $I$-$V$ trace. (d) Estimated PbTe-Pb interface transparency. Red and blue dots are based on the $I_{\text{excess}}$ and $R_{\text{n}}$ extracted from the positive and negative bias branches, respectively.  (e) $R$ versus $I$ and $B$. $R=$d$V/$d$I$. $B$ is parallel to the nanowire. (f) Line cuts from (e). (g) $R$ versus $I$ and $B$. $B$ is perpendicular to the substrate. (h) Line cuts from (g). (i) $T$ dependence.  }
\label{fig2}
\end{figure}

Figure 2(a) shows the $I$-$V$ characteristic of the JJ as a function of gate voltage ($V_{\text{G}}$). The ``white triangle''  is the gate-tunable supercurrent region, see Fig. 2(b) for line cuts. A series resistance ($R_{\text{series}}$), including the filters ($\sim$3.5 k$\Omega$) and contacting resistance ($\sim$400 $\Omega$), is subtracted from the raw two-terminal $I$-$V$ curve. See Fig. S1 in SM for details. The current sweeps from negative to positive. The current bias at which the JJ jumps from the superconducting to the resistive branch defines the switching current $I_{\text{s}}$. $I_{\text{s}}$ of 78 nA translates to a Josephson energy $E_{\text{J}}=\hbar I_{\text{c}}/2e \sim$160 $\mu$eV, well exceeding the typical $E_{\text{J}}$ (90 $\mu$eV)  for a gatemon \cite{2015_PRL_gatemon}. We expect the critical current $I_{\text{c}}$ to be close to $I_{\text{s}}$ since the fridge $T$ is much less than $E_{\text{J}}$ ($\sim$1.8 K). The ``switching point'' from the resistive to the superconducting branch (the negative current bias) defines the retrapping current $I_{\text{r}}$. $I_{\text{r}}$ being almost identical to $I_{\text{s}}$ suggests that the JJ is in the over-damped regime. 

To study the PbTe-Pb interface transparency, Fig. 2(c) shows an $I$-$V$ curve over a larger $I$ range. The linear fit (red dashed line) for $V>2\Delta$ extracts the JJ normal state resistance, $R_{\text{n}}\sim$941 $\Omega$. $\Delta$ is the size of the superconducting gap (see Fig. 3). Extending the red dashed line to the $I$ axis gives the excess current, $I_{\text{excess}}\sim$307 nA. We then calculate $eI_{\text{excess}}R_{\text{n}}/\Delta \sim$0.69, which can be further used to estimate the junction transparency ($\sim$0.65) \cite{BTK, Flensberg_1988}. Figure 2(d) shows the transparency as a function of $V_{\text{G}}$. The transparency (equals to $1/(1+Z^2)$) is calculated by solving the equation of $Z$: $eI_{\text{excess}}R_{\text{n}}/\Delta = 2(1+2Z^2)\times \text{tanh}^{-1}[2Z\sqrt{(1+Z^2)/(1+6Z^2+4Z^4)}] \times [Z\sqrt{(1+Z^2)(1+6Z^2+4Z^4)}]^{-1} - 4/3$ \cite{Niebler_2009}. The red (blue) dots used $I_{\text{excess}}$ and $R_{\text{n}}$ extracted from the positive (negative) bias axis in Fig. 2(c). The maximum transparency exceeding 0.8 indicates a high quality PbTe-Pb interface. Note that the extracted transparency here is probably underestimated since  $\Delta$ of 0.42 meV (in the tunneling regime) is used. $\Delta$ in the open regime is smaller (see Fig. 3). For detailed analysis of $I_{\text{excess}}$, $I_{\text{s}}$ and $R_{\text{n}}$, see Fig. S2 in SM.

Figures 2(e) and 2(f) show the magnetic field ($B$) dependence of the supercurrent. $B$ is roughly parallel to the nanowire axis. For clarity, the differential resistance, $R=$d$V/$d$I$, is presented. $I_{\text{s}}$ decreases monotonically, suggesting that the orbital effect in PbTe does not play a significant role. In Fig. S3 in SM, we show another device which resolves clear supercurrent interference possibly due to the orbital effect in the PbTe nanowire. In Figs. 2(g) and 2(h), $B$ is aligned perpendicular to the substrate (Pb film) and the supercurrent is suppressed at $\sim$0.07 T. This field value is roughly consistent with the critical field (0.08 T) of a bulk superconductor Pb. For $B$ parallel to the nanowire (Fig. 2(e)), the critical field is two times larger, indicating less orbital effect in the Pb film (the effective Pb area perpendicular to $B$ is smaller due to the thin thickness of the film). The multiple peaks are subgap features ($V<2\Delta/e$) likely result from multiple Andreev reflections (MARs) \cite{MAR}. For completeness, we show the $T$ dependence of the supercurrent in Fig. 2(i). The supercurrent is fully suppressed at $T \sim$1 K, much smaller than the critical $T$ of a bulk Pb ($\sim$7 K).

Driving $V_{\text{G}}$ more negative lowers the junction transmission and reaches the tunneling regime. The differential conductance, $\text{d}I/\text{d}V$, in this regime can resolve the superconducting gap as shown in Fig. 3(a). The two peaks, symmetrically located at $V \sim \pm$0.84 mV, correspond to the 2$\Delta$ coherence peaks, see Fig. 3(b) for the line cut. The negative differential conductance outside the gap (next to the coherence peak) is typical for S-NW-S devices (S for superconductor and NW for nanowire) \cite{Gul2017}. The subgap conductance reaches zero (close to the measurement noise level), suggesting a hard superconducting gap. The ratio of outside gap versus subgap conductance is close to two orders of magnitude. Note that for S-NW-S devices, the tunneling conductance reflects the convolution of two density of states (DOS) of superconductor quasiparticles. To directly reveal the DOS, an N-NW-S device (N for normal metal) is more appropriate. More positive $V_{\text{G}}$ in Fig. 3(a) reveals a sharp zero-bias conductance peak, resulting from the supercurrent (in the open regime). Several subgap peaks are also visible in Fig. 3(a), possibly due to MARs \cite{MAR} (see Fig. S2 in SM for additional analysis).

\begin{figure}[tb]
\includegraphics[width=\columnwidth]{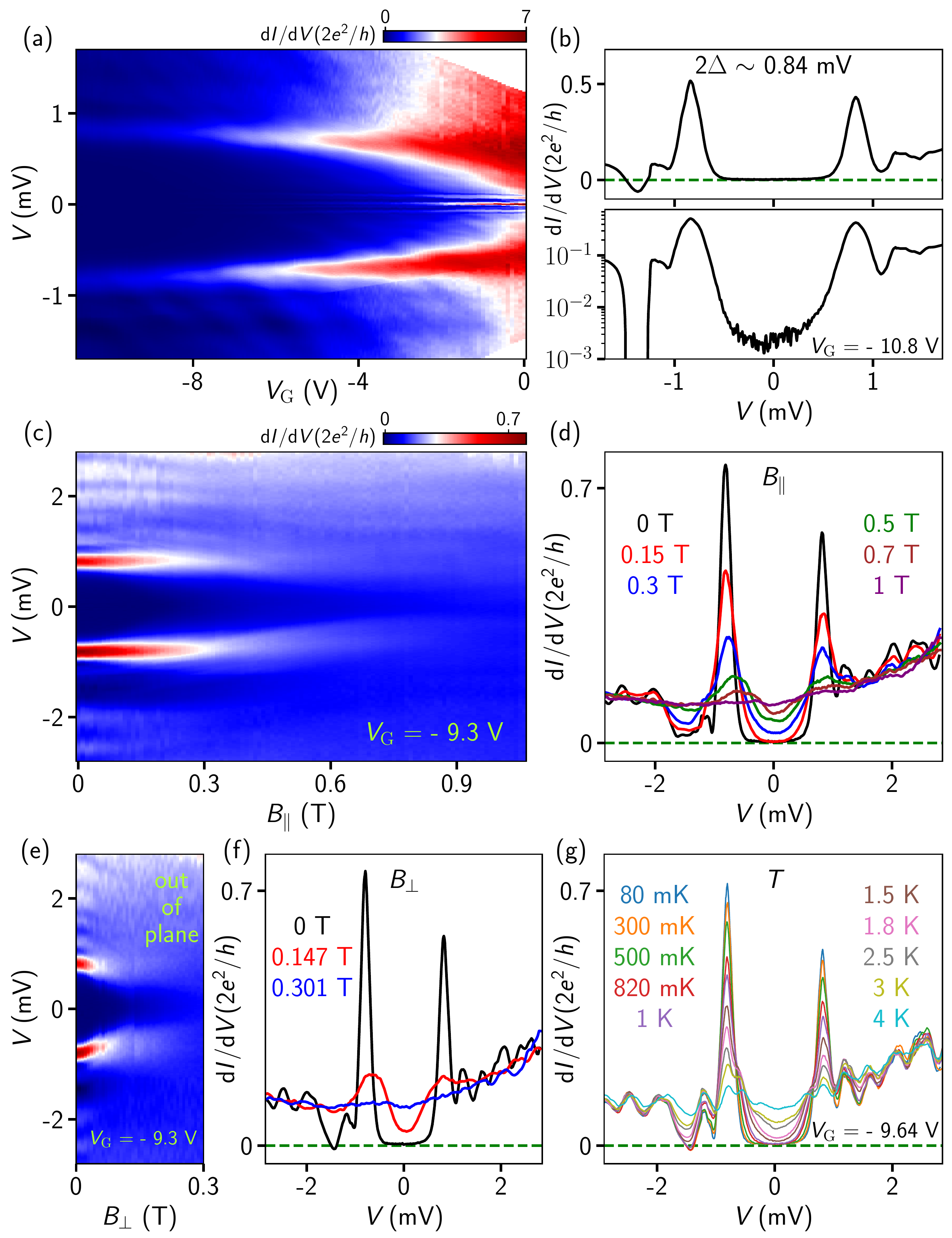}
\centering
\caption{Superconducting gap in the tunneling regime of Device A. (a) $\text{d}I/\text{d}V$ versus $V$ for more negative $V_{\text{G}}$. $B$ = 0 T. (b) A line cut in the tunneling regime, plotted in linear (upper) and logarithmic (lower) scales. (c) $B$ (parallel to the nanowire) dependence of the gap. (d) Line cuts from (c). (e) $B$ is perpendicular to the substrate. (f) Line cuts from (e). (g) $T$ dependence of the gap. }
\label{fig3}
\end{figure}

The size of the induced gap $\Delta \sim$0.42 meV is significantly smaller than the gap of a bulk Pb, estimated based on the formula $\Delta=1.76k_{\text{B}}T_{\text{c}} \sim$1.1 meV  ($T_{\text{c}} \sim$7 K for Pb). In a recent work on InAs epitaxially coupled to Pb \cite{InAs-Pb}, an induced gap of 1.1-1.5 meV is reported with the Pb film thickness being 9-50 nm. As a comparison, our Pb film thickness is $\sim$23 nm. The relatively small induced gap suggests that our PbTe-Pb is likely not in the strong coupling regime, but rather an intermediate coupling case. The advantages and disadvantages for strong and intermediate couplings have been extensively studied in theory \cite{LutchynSchrodinger,Flensberg_SP, DasSarma_SP, Loss_metalization}. Most of the current hybrid nanowires (InAs-Al, InAs-Pb and InSb-Al) are in the strong coupling regime. Our device provides a platform to study the proximity effect in the intermediate coupling regime, shedding light on e.g., the role of disorder in the superconductor \cite{DasSarma_disorder_superconductor_2016, DasSarma_disorder_superconductor_2022}.

Another notable feature in Fig. 3(a) is the variation of the gap size: The gap ``shrinks'' for a more positive $V_{\text{G}}$. We have independently calibrated the fridge filters and confirm that this shrink is not due to an overestimation of $R_{\text{series}}$ and its shared bias voltage. A possible explanation for the shrink is the gate-tunable superconductor-semiconductor coupling \cite{Michiel2018}. More positive $V_{\text{G}}$ ``drags'' the electron wavefunctions more into the PbTe (less in Pb), leading to a smaller induced gap. Though the side gate has a ``finger'' shape, the large dielectric constant of PbTe can bend the electric field lines such that $V_{\text{G}}$ can also tune the proximitized PbTe parts \cite{Fabrizio_PbTe}.  

Figures 3(c)-(f) show the $B$ dependence of the gap, resolving a critical field of $\sim$0.9 T for parallel $B$ and $\sim$0.3 T for perpendicular $B$. The critical field of a bulk Pb is 0.08 T. The differences between these three values indicate the reduced orbital effect in Pb for smaller sizes. To achieve higher critical field (a few Tesla), a thinner ($\sim$10 nm) Pb film is needed in future devices \cite{InAs-Pb}.

The small oscillations outside the gap in Figs. 3(d) and 3(f) (also visible in Fig. 3(a)) are reminiscent of those in Ref. \cite{Morten_doubling}. Since the PbTe-Pb interface is rather flat, we think the oscillations are likely caused by the nonuniform PbTe-CdTe interface. Though the substrate is typically flat after the growth of the CdTe buffer, PbTe growth requires heating the substrate to 318.5 $^{\circ}$C to have selectivity. This temperature can evaporate part of the CdTe buffer, causing the non-flat interface (see Fig. S4 in SM). Electron back scattering can occur, leading to those Fabry-Perot-like oscillations. This non-uniformity surely degrades the device quality and should be minimized in future optimizations. In Fig. S3, we show two devices which can not be pinched off before the gate leaks, possibly due to this substrate disorder.

\begin{figure}[tb]
\includegraphics[width=\columnwidth]{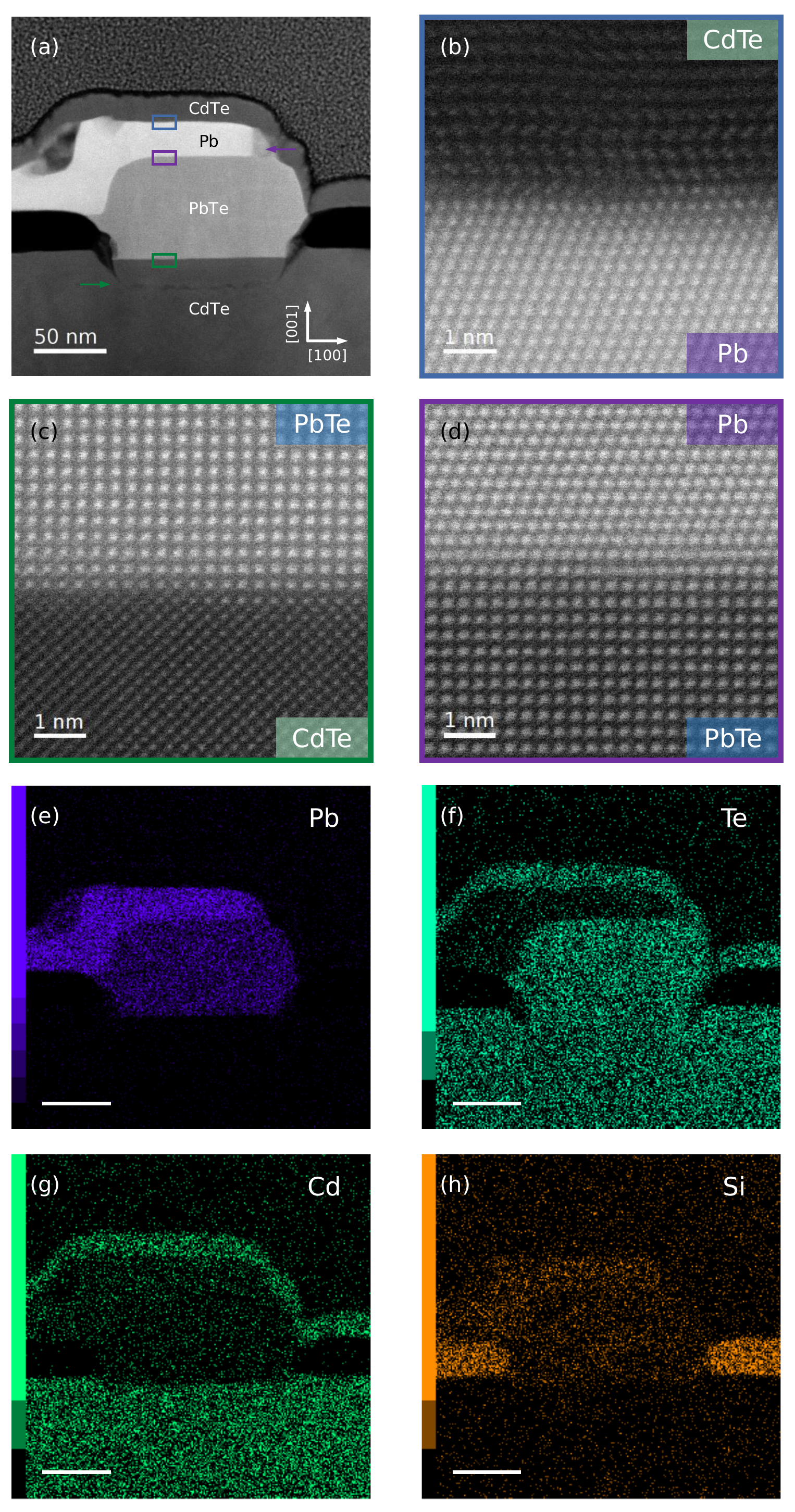}
\centering
\caption{STEM analysis of device B. (a) HAADF image of the nanowire cross-section with crystal directions labeled. (b) Atomically resolved image near the Pb-CdTe (capping) interface, see the blue box in (a). (c) Image of the PbTe-CdTe interface, see the green box in (a). (d) Image of the PbTe-Pb interface, see the violet box in (a). (e)-(f), EDX maps of Pb, Te, Cd and Si, respectively. Scale bar, 50 nm. }
\label{fig4}
\end{figure}

Figure 3(g) shows the $T$ dependence of the gap. The gap is completely ``washed out'' at $\sim$4 K. The small oscillations, however, are still present. This is consistent with our interpretation above: The normal scattering (Fabry-Perot-like) can have a different energy scale than the superconducting correlations. In Figs. 3(d) and 3(f), the oscillations can be suppressed, possibly due to the accumulated phases induced by $B$.

After the quantum transport characterization, we cut the device cross-section using a focused ion beam and performed the high-resolution scanning transmission electron microscopy (STEM). Figure 4(a) shows the image of the cross-section lamella from another device (device B). The lamella of device A was damaged by the focused ion milling (see Fig. S4 in SM). Devices A and B were grown and fabricated together (on one same substrate chip). The transport study of device B is shown in Fig. S3 in SM. The high-angle annular dark field (HAADF) image in Fig. 4(a) clearly shows different material layers with sharp interfaces. The green arrow points to a dark line region in the CdTe(001), likely the residue of Ar treatment during the substrate cleaning. By growing the CdTe buffer before PbTe, we can ``bury'' this disordered region in the CdTe, away from the PbTe nanowire. We find that this buffer growth can significantly improve the device transport quality compared to our previous study \cite{Jiangyuying}. The violet arrow indicates a corner of the Pb film which was damaged during the lamella preparation. For additional STEM analysis on device B, see Fig. S5 in SM. 

Figures 4(b)-(d) show the atomically resolved STEM images at three key interfaces: the Pb-CdTe (capping), the PbTe-CdTe (substrate buffering) and the PbTe-Pb. The PbTe and CdTe are lattice matched (Fig. 4(c)). The PbTe-Pb interface is also sharp but with mismatched dislocations. This disorder could be reduced in the future by growing a thin layer of CdTe between the PbTe and the Pb growth. The CdTe capping in Fig. 4(b) prevents the oxidation of the Pb film underneath. The lattice mismatch between Pb and CdTe is a disorder source in the superconductor. This disorder may enhance the proximity effect \cite{LutchynSchrodinger} and its disadvantage needs further systematic studies. The energy-dispersive x-ray spectroscopy (EDX) maps of the four elements Pb, Te, Cd and Si (for the dielectric mask) are shown in Figs. 4(e)-(h). Both the STEM and the EDX suggest little interlayer diffusion. 

To summarize, we have studied the superconducting proximity effect in PbTe-Pb hybrid nanowires. A Josephson junction device based on this hybrid reveals a gate-tunable supercurrent in the open regime and a hard induced superconducting gap in the tunneling regime. Though the device is far from perfect and more optimizations are still needed (e.g. substrate uniformity), our work shows that the combination of PbTe and Pb could work as a new hybrid nanowire system for Majorana searches.

\textbf{Acknowledgment} This work is supported by Tsinghua University Initiative Scientific Research Program, National Natural Science Foundation of China (92065206) and the Innovation Program for Quantum Science and Technology (2021ZD0302400). Raw data and processing codes within this paper are available at https://doi.org/10.5281/zenodo.7398019.

\bibliography{mybibfile}% Produces the bibliography via BibTeX.

\newpage

\onecolumngrid

\newpage


\section*{Supplementary material}

# Supplementary Information for "Proximity effect in PbTe-Pb hybrid nanowire Josephson junctions"

Zitong Zhang,[1, *] Wenyu Song,[1, *] Yichun Gao,[1, *] Yuhao Wang,[1, *] Zehao Yu,[1] Shuai Yang,[2] Yuying Jiang,[1] Wentao Miao,[1] Ruidong Li,[1] Fangting Chen,[1] Zuhan Geng,[1] Qinghua Zhang,[3] Fanqi Meng,[4] Ting Lin,[3] Lin Gu,[4] Kejing Zhu,[2] Yunyi Zang,[2] Lin Li,[2] Runan Shang,[2] Xiao Feng,[1, 2, 5] Qi-Kun Xue,[1, 2, 5, 6] Ke He,[1, 2, 5, †] and Hao Zhang[1, 2, 5, ‡]

[1]*State Key Laboratory of Low Dimensional Quantum Physics,
Department of Physics, Tsinghua University, Beijing 100084, China*
[2]*Beijing Academy of Quantum Information Sciences, Beijing 100193, China*
[3]*Institute of Physics, Chinese Academy of Sciences, Beijing 100190, China*
[4]*School of Materials Science and Engineering, Tsinghua University, Beijing 100084, China*
[5]*Frontier Science Center for Quantum Information, Beijing 100084, China*
[6]*Southern University of Science and Technology, Shenzhen 518055, China*

## Method

### Substrate fabrication

Shadow-walls and markers. Hydrogen SilsesQuioxane (FOx-16, Dow Corning), the HSQ resist, was spun onto the CdTe (001) substrate at 2000 rpm for 60 s and then baked at 150 °C for 60 s. An electron beam lithography (EBL) was then performed to write the shadow-wall and marker patterns. The chip was immersed in tetramethylammonium hydroxide (TMAH) for 180 s to dissolve the unexposed resist regions.

$Si_xN_y$ mask. The chip was then covered by a thin film of $Si_xN_y$ (30-40 nm) using the plasma enhanced chemical vapor deposition (PECVD). Then AR-P 672.045 resist was spun at 4000 rpm for 60 s and baked at 130 °C for 10 min. Another EBL was performed to write the designed nanowire trenches and the markers (for alignement during the device fabrication). The chip was then developed (MIBK:IPA=1:3) for 40 s. Reactive ion etching ($CHF_3$ with $O_2$) was used to etch away the trenched regions of the $Si_xN_y$ film. The resist was removed in acetone and the substrate was cleaned in oxygen plasma for 8 min at a power of 60 W. The CdTe chip covered by a $Si_xN_y$ mask was ready for the next stage growth.

### Selective area growth of PbTe nanowires and Pb deposition

PbTe nanowires. The chip was loaded into an ultra-high vacuum molecular beam epitaxy (MBE) chamber (base pressure less than $2.0 \times 10^{-10}$ mbar). To remove the native oxide layer of CdTe, Ar bombardment was performed at a beam energy of 0.9 KeV and a beam current density of 0.7 $\mu Acm^{-2}$. The chip was then annealed at ~250 °C for 50 min. A CdTe buffer layer was grown at ~270 °C with a beam flux of ~10 nm/h. Then, PbTe nanowires were grown at ~320 °C with a beam flux of ~0.6 nm/min under Te atmosphere.

Pb deposition. The Pb film was in situ deposited at another sample stage in the same MBE chamber. The substrate was cooled to ~100 K with liquid nitrogen during the deposition. Finally, a CdTe layer (10-15 nm) was grown to cap the entire chip.

### Device fabrication

Pb Etching. To prevent a short circuit between the source and the drain electrodes, most of the Pb film on the substrate needs to be etched away. First, the device regions were manually coated by the AR-P 671.05 resist using a toothpick. Pb film on the edges was then etched in an ATC-IM Ion Milling System (Ar ion milled for 260 s). The resist was removed in acetone. The chip was coated by a bi-layer of AR-P 671.05 at 4000 rpm (~490 nm thick for each layer). Then the resist was vacuum pumped (no hot plate baking) followed by an EBL to write the etching patterns. After developing in diluted MIBK for 50 s and IPA for 30 s, a second ion-milling was performed to remove the Pb film on the substrate near the nanowire. The resist was removed in acetone.

Contacts and gate. The standard EBL was performed to evaporate Ti/Au (10/65 nm). Resist, 490-nm-thick AR-P 671.05. Before evaporation, argon etching (160 s, 50 W, $5.4 \times 10^{-2}$ Torr) was performed in the load-lock to remove the CdTe capping in the contact regions. Lift-off in acetone overnight, followed by IPA and nitrogen gun blow-dry.

---

* equal contribution
† kehe@tsinghua.edu.cn
‡ hzquantum@mail.tsinghua.edu.cn

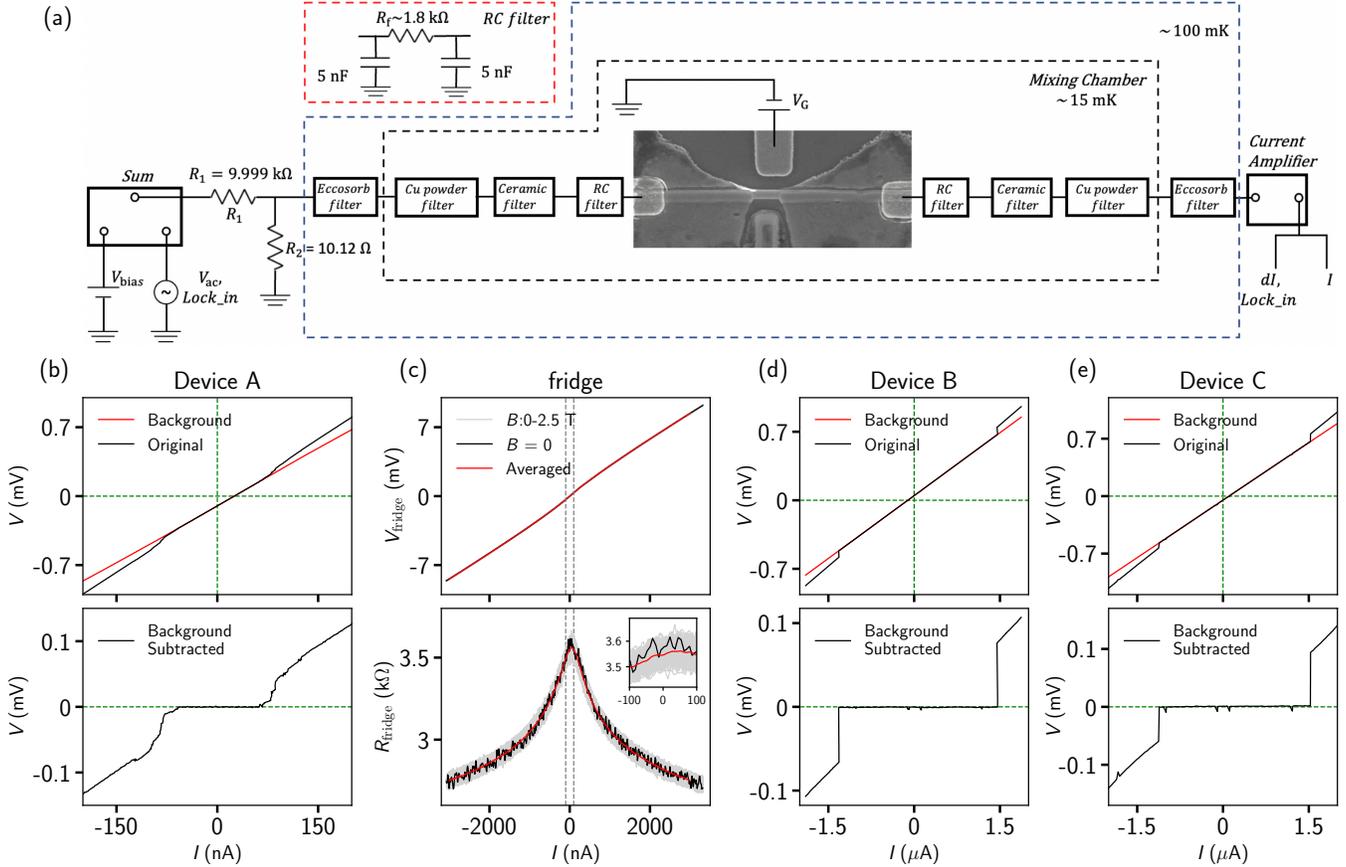

Fig. S 1. (a) Measurement circuit diagram. A DC voltage ($V_{bias}$) was first summed with a lock-in excitation ($V_{ac}$), and then passed through a voltage divider ($R_1$ vs $R_2$). The ratio of the voltage before versus after the divider is calibrated to be 1011 (including the output resistance of the summing module) with an uncertainty of 0.5%. After the four-stage filters, this voltage was applied to the source electrode. The current was measured from the drain contact. For clarity, the four-stage filters were not drawn for the fridge line of the gate. (b) Upper panel, the black curve is the measured $I$-$V$ including the contribution of $R_{series}$. $V$ here refers to the bias voltage after the voltage divider (before the filters). The jumps on the black curve indicate the presence of supercurrent ($I_s$ and $I_r$). The red curve is the background fit contributed by $R_{series}$. A small vertical shift (98 $\mu$V) is corrected to account for the equipment offset so that the curve passes through the origin of coordinates. $R_{series}$ includes the device contact resistance ($\sim$400 $\Omega$) and $R_{fridge}$. $R_{fridge}$ includes the filters of two fridge lines, the fridge lines, the output resistance of the divider and the input resistance of the current amplifier. The later three components are small. Lower panel, $I$-$V$ of the JJ after subtracting the background. $I_s$ and $I_r$ were defined where $|V| = 10$ $\mu$V. (c) $I$-$V$ (upper) and differential resistance (lower) characterizations of $R_{fridge}$. The two fridge lines of device A were shorted at the printed circuit board (PCB) in a separate cool-down. $T \sim$15 mK. $R_{fridge}$ shows little dependence on $B$ (black for 0 T, gray for finite $B$ and red the average). The red curves are used to calibrate the background in (b). $R_{fridge}$ (mainly the filters) varies when the current bias is large. This variation mainly affects the estimation of $R_n$. Within the supercurrent branch (below 100 nA, the vertical dashed lines), variation of $R_{fridge}$ is small (see inset). (d) Upper panel, an example of the original $I$-$V$ trace (black) and the background fit (red) for device B. Note that the measurement circuit for device B (and C) is "pseudo-four-terminal": two fridge lines were bonded to the source contact (and also the drain). One to bias the current and one to measure the voltage. The background excludes $R_{fridge}$ and is only the contribution of contact resistance (428 $\Omega$) of device B (plus a 48 $\mu$V vertical shift due to the equipment offset). Lower panel, $I$-$V$ of the JJ after subtracting the background. (e) Same with (d) but for device C. The contact resistance is 467 $\Omega$ and the voltage offset is 49 $\mu$V.

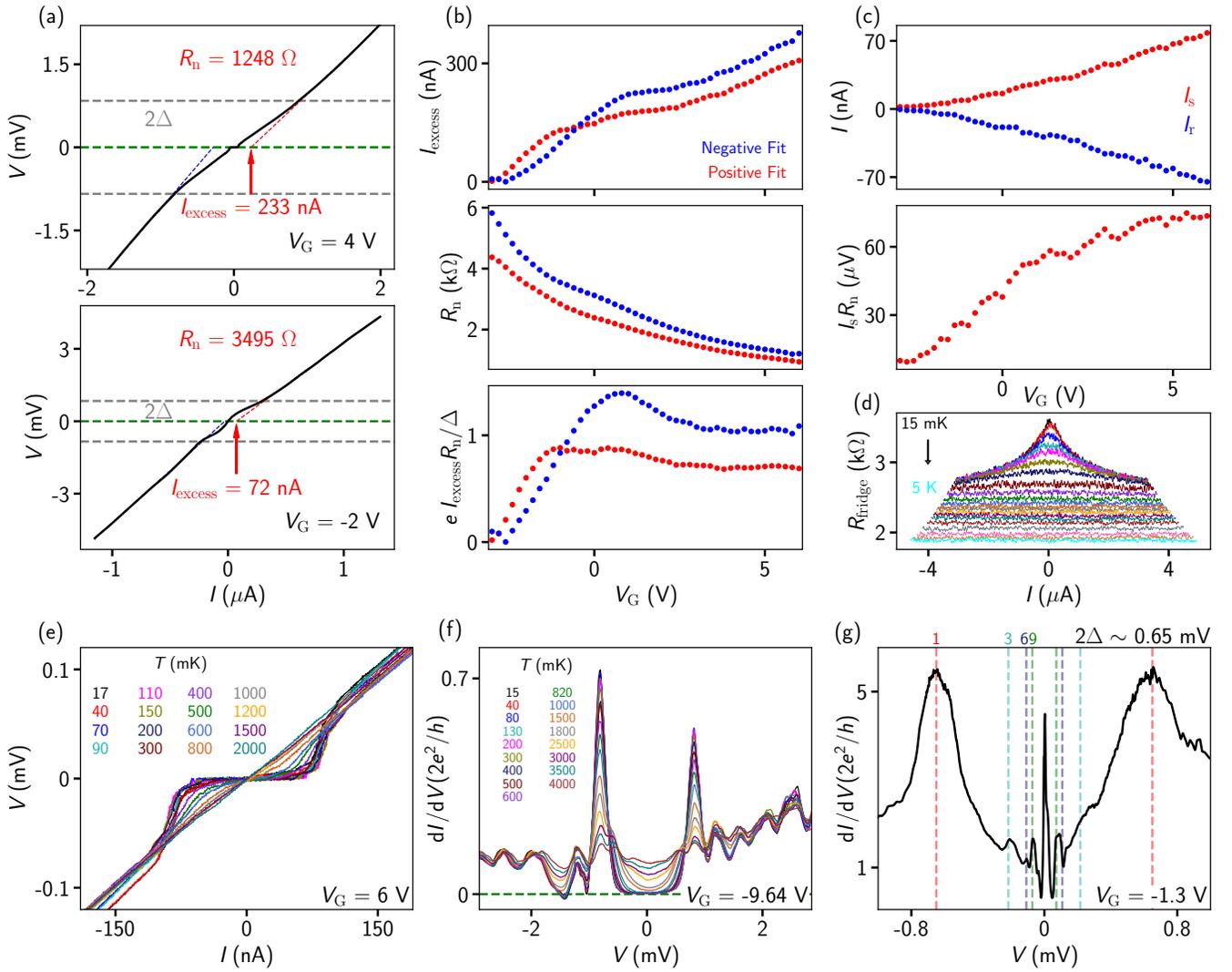

Fig. S 2. (a) Two additional $I$-$V$ examples for the estimations of $I_{\text{excess}}$ and $R_{\text{n}}$. (b) Extracted $I_{\text{excess}}$, $R_{\text{n}}$ and $eI_{\text{excess}}R_{\text{n}}/\Delta$ as a function of $V_{\text{G}}$. (c) Extracted $I_{\text{s}}$, $I_{\text{r}}$ and $I_{\text{s}}R_{\text{n}}$ as a function of $V_{\text{G}}$. (d) $T$ dependence of $R_{\text{fridge}}$ for device A, used to to calibrate the background subtraction in the $T$ dependence (Fig. 2(i) and Fig. 3(g)). (e) $T$ dependence of the supercurrent (Fig. 2(i)) including all the $T$'s. The contact resistance may vary over $T$ and a value of 300 $\Omega$ is used. (f) $T$ dependence of the gap (Fig. 3(g)) including all the $T$'s. (g) A $dI/dV$ line cut of Fig. 3(a) in the open regime. The zero-bias conductance peak is due to supercurrent. The subgap peaks are likely MARs at bias positions $V_N = 2\Delta/N$, with $N$ being an integer number (see the labeling and dashed lines for guidance). Note that $\Delta$ (0.325 meV) in this open regime is smaller than that in the tunneling regime (0.42 meV).

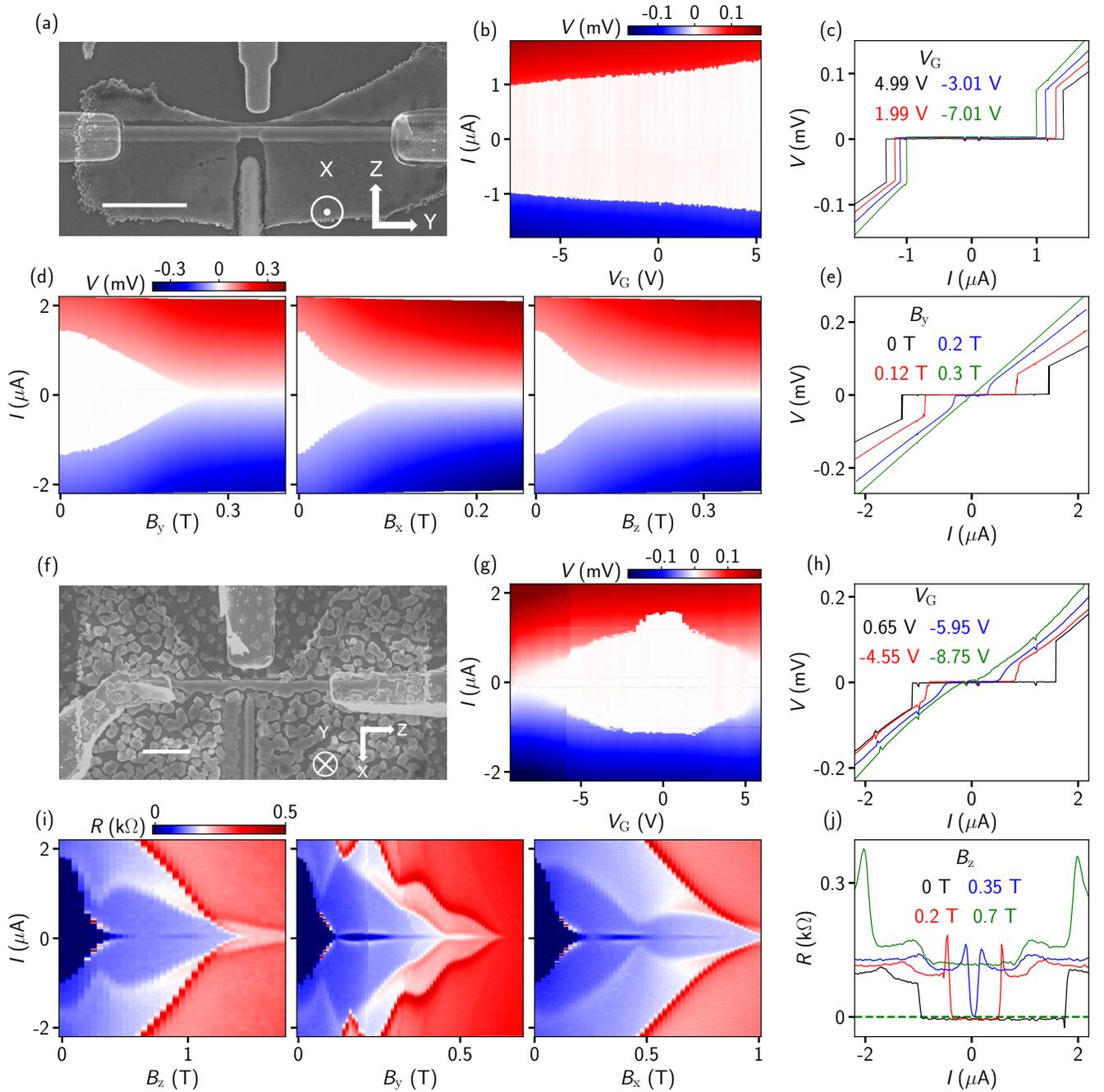

Fig. S 3. Two additional PbTe-Pb hybrid devices. (a) SEM of device B whose cross-sectional STEM is shown in Fig. 4. Scale bar, 1 $\mu$m. (b) Gate-tunable supercurrent in device B. The gate starts to leak beyond -7 V, therefore the tunneling regime cannot be reached. (c) Several line cuts from (b). (d) $B$ dependence of the supercurrent along three different directions (see the labeling in (a)). (e) Line cuts from (d) for $B_y$. (f) SEM of device C. Scale bar, 1 $\mu$m. The small irregular "islands" on the substrate are Pb (capped by CdTe). The Pb film for device C was designed to be thin and it became discontinuous on the substrate. Device C was grown separately from devices A and B. (g) Gate-tunable supercurrent in device C. This device also cannot reach the tunneling regime before the gate leaks. (h) Line cuts from (g). $I_r$ is less than $I_s$, suggesting that the JJ is in the under-damped regime or purely a heating effect. (i) $B$ dependence of the supercurrent (shown in differential resistance) and subgap features. The interference pattern suggests possible orbital effect in the nanowire. (j) Line cuts from (i) for $B_z$. The zero resistance region is actually below zero by 5 $\Omega$, possibly due to an overestimation of $R_{series}$. The $R_{series}$ of 467 $\Omega$ is a fit of the DC data while for lock-in, there may be a slight difference due to the finite cut-off frequency of the amplifiers.

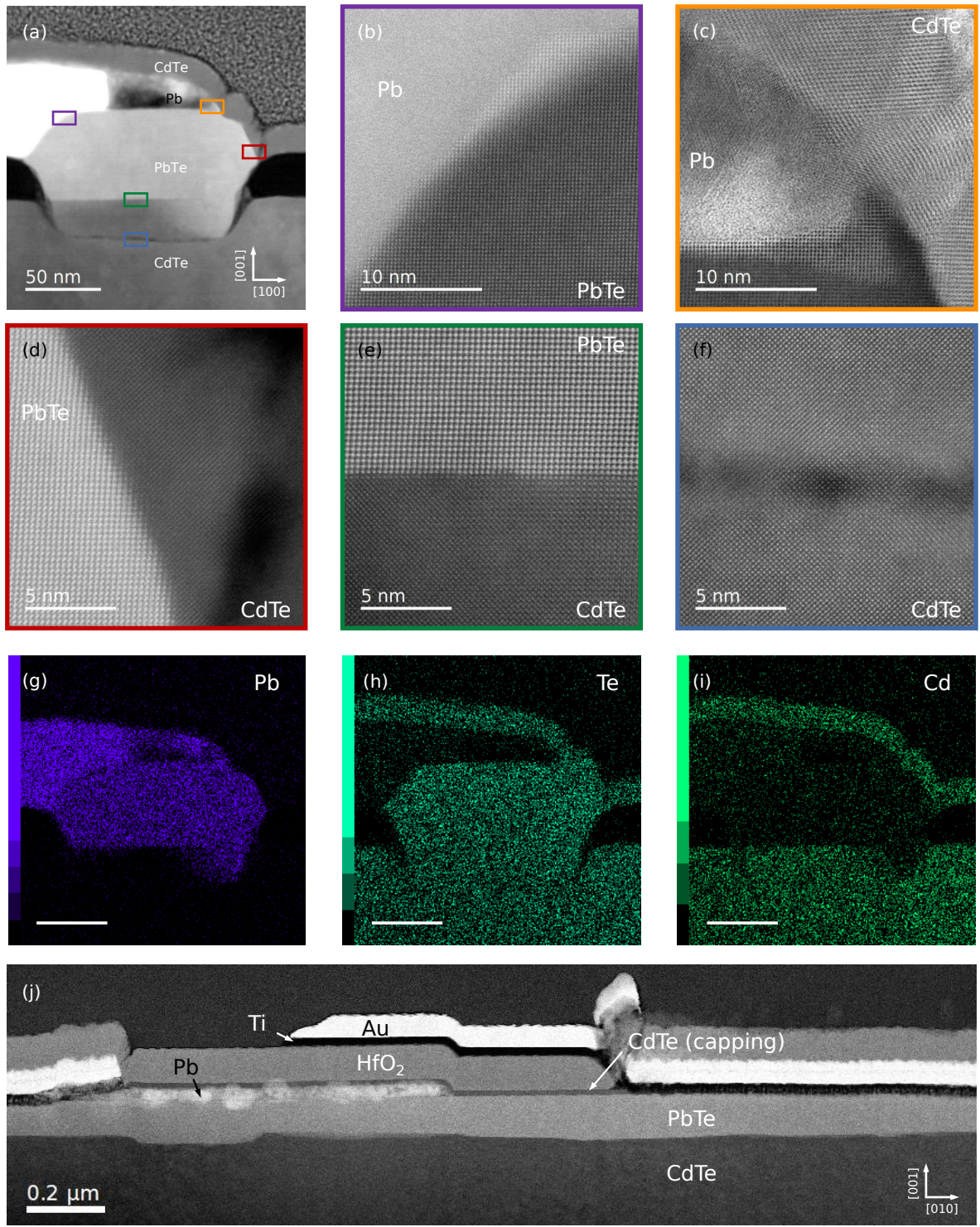

Fig. S 4. STEM results of device A (for panels (a)-(i)). (a) Cross-section image of device A. The Pb film was damaged by the focused ion beam during the lamella preparation. (b)-(f), Zoom-in of the interfaces in (a) (see the corresponding colored boxes). The dark line in the middle of (f) is the residue of the substrate cleaning (Ar treatment and annealing). (g)-(i) EDX maps of the device. Scale bar, 50 nm. The bottom right corner of PbTe is due to the partial evaporation of the CdTe buffer during the PbTe growth and should be minimized in the future optimizations. (j) STEM image cut along the nanowire axis from another PbTe-Pb nanowire. Note that this wire was grown under the similar condition with devices A and B. The Pb film was damaged during the lamella preparation. The non-uniform PbTe-CdTe interface is likely due to the evaporation of CdTe during the PbTe growth. This interface disorder is probably the cause of the Fabry-Perot-like oscillations in Fig. 3 and should be eliminated in future optimizations.

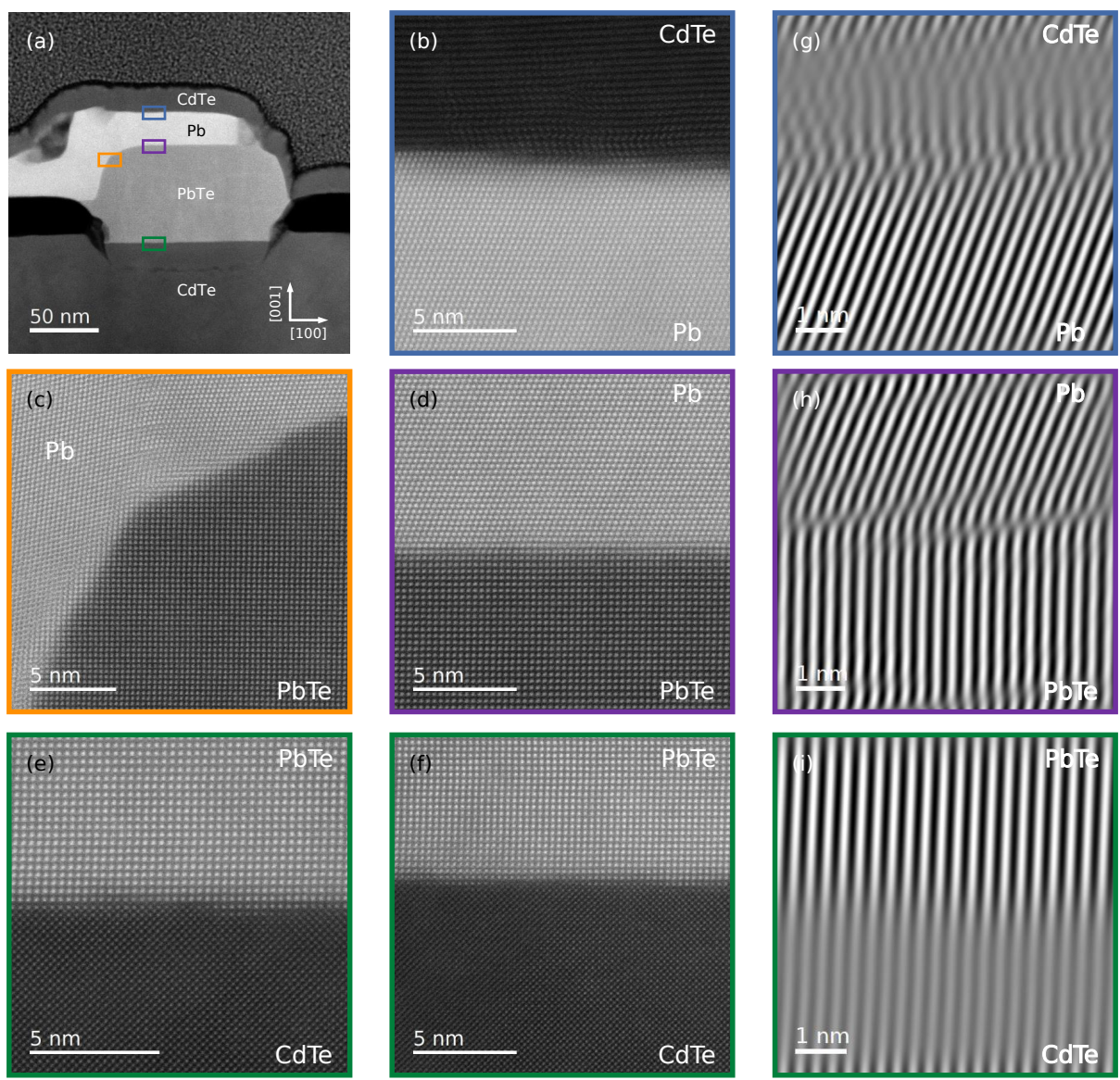

Fig. S 5. Additional STEM results of device B. (a) Cross-section image of device B. (b)-(f) Atomically resolved images at the interfaces (see panel (a) for the corresponding colored boxes). (g), (h) and (i) are the Inverse fast Fourier transformation (IFFT) of Figs. 4(b), 4(d) and 4(c). IFFT can resolve the columns of atoms as vertical lines. The dislocations in Pb-CdTe and Pb-PbTe interfaces are due to lattice mismatch. The PbTe-CdTe interface shows good lattice matching.



\end{document}